\documentclass[english]{article}
\usepackage[T1]{fontenc}
\usepackage{a4}
\usepackage{amsmath}
\usepackage{babel}
\usepackage{color}
\usepackage{graphics}
\usepackage{setspace}
\makeatletter

\newcommand{\LyX}{L\kern-.1667em\lower.25em\hbox{Y}\kern-.125emX\spacefactor1000}

\makeatother

\begin{document}

\title{Emission time scale of light particles in the system Xe+Sn at 50 AMeV. A probe
for dynamical emission ?}

\maketitle
{\centering D.~Gourio\( ^{1} \), D.~Ardouin\( ^{2} \), M.~Assenard\( ^{2} \),
G.~Auger\( ^{3} \), Ch.O.~Bacri\( ^{4} \), N.~Bellaize\( ^{5} \), A.~Benkirane\( ^{3} \),
J.~Benlliure\( ^{3} \), B.~Berthier\( ^{6} \), E.~Bisquer\( ^{7} \), B.~Borderie\( ^{4} \),
R.~Bougault\( ^{5} \), P.~Box\( ^{4} \), R.~Brou\( ^{5} \), J.L.~Charvet\( ^{6} \),
A.~Chbihi\( ^{3} \), J.~Colin\( ^{5} \), D.~Cussol\( ^{5} \), R.~Dayras\( ^{6} \),
E.~De~Filippo\( ^{6} \), A.~Demeyer\( ^{7} \), C.~Donnet\( ^{7} \), D.~Durand\( ^{5} \),
P.~Ecomard\( ^{3} \), P.~Eudes\( ^{2} \), M.~Germain\( ^{2} \), D.~Guinet\( ^{7} \),
L.~Lakehal-Ayat\( ^{4} \), P.~Lautesse\( ^{7} \), J.L.~Laville\( ^{3} \),
L.~Lebreton\( ^{7} \), C.~Le~Brun\( ^{5} \), J.F.~Lecolley\( ^{5} \), T.~Lefort\( ^{5} \),
A.~Lefèvre\( ^{3} \), R.~Legrain\( ^{6} \), N.~Le~Neindre\( ^{5} \), O.~Lopez\( ^{5} \),
M.~Louvel\( ^{5} \), N.~Marie\( ^{3} \), V.~Métivier\( ^{2} \), L.~Nalpas\( ^{6} \),
A.~Ouatizerga\( ^{4} \), M.~Parlog\( ^{8} \), J.~Péter\( ^{5} \), E.~Plagnol\( ^{4} \),
E.~Pollaco\( ^{6} \), A.~Rahmani\( ^{2} \), R.Régimbart\( ^{5} \), T.~Reposeur\( ^{2} \),
M.F.~Rivet\( ^{4} \), E.~Rosato\( ^{9} \), F.~Saint-Laurent\( ^{3} \), S.~Salou\( ^{3} \),
M.~Squalli\( ^{4} \), J.C.~Steckmeyer\( ^{5} \), G.~Tabacaru\( ^{8} \), B.~Tamain\( ^{5} \),
L.~Tassan-Got\( ^{4} \), E.~Vient\( ^{5} \), C.~Volant\( ^{6} \) J.P.~Wieleczko\( ^{3} \),
A.~Wieloch\( ^{5} \) and K.~Yuasa-Nakagawa\( ^{5} \).\par}
\bigskip{}

{\small \( ^{1} \)GSI mbH, D-64291 Darmstadt, Germany.}{\small \par}

{\small \( ^{2} \)SUBATECH, IN2P3-CNRS et Université, F-44070 Nantes Cedex,
France.}{\small \par}

{\small \( ^{3} \)GANIL, CEA et IN2P3-CNRS, B.P.~5027, F-14076 Caen Cedex,
France.}{\small \par}

{\small \( ^{4} \)IPN, IN2P3-CNRS, F-91406 Orsay Cedex, France.}{\small \par}

{\small \( ^{5} \)LPC, IN2P3-CNRS, ISMRA et Université, F-14050 Caen Cedex,
France.}{\small \par}

{\small \( ^{6} \)DAPNIA/SPhN, CEA/Saclay, F-91191 Gif sur Yvette Cedex, France.}{\small \par}

{\small \( ^{7} \)IPN, IN2P3-CNRS et Université, F-69622 Villeurbanne Cedex,
France.}{\small \par}

\textcolor{black}{\small \( ^{8} \)}\textcolor{black}{\small NIPNE, RO-76900 Bucharest-M\u{a}gurele, Romania.}{\small \par}

{\small \( ^{9} \)DSFSezione INFN, Università di Napoli ``Federico II'', I-80126
Napoli, Italy.}{\small \par}
\bigskip{}

\begin{abstract}
Proton and deuteron correlation functions have been investigated with both impact
parameter and emission source selections. The correlations of the system \( ^{129}Xe+^{Nat}Sn \)
at 50 AMeV have been measured with the 4\( \pi  \) INDRA which provides a complete
kinematical description of each event. The emission time scale analyzed with
a quantum model reveals the time sequence of the light particles emitted by
the projectile-like fragment. The short and constant emission time of the proton,
independent of the impact parameter, can be attributed to a preequilibrium process.
\end{abstract}

\section{Introduction }

The microscopic description of hot nuclear matter is fundamental to the understanding
of energetic heavy ion collisions. Many of the theoretical models developed
so far suffer from a shortage of observables to directly test the properties
of nuclear matter early after the reaction. An example is the time scale of
the reaction. By the very nature, experimenters observe only the signals reaching
the detectors at infinite time. Subsequently only asymptotic properties of particles
leaving the reaction zone are monitored. This shortcoming is partly bypassed
in studies using interferometry of light particles \cite{Koonin}-\cite{Review_Corinne}.
It has been shown that the two-body correlation function is not only sensitive
to emission time and source size, but also to charge and slope parameter of
the source \cite{Martin_3Body} which affect the velocity distribution of the
particles and thus the relative distance between them.

It is quite obvious that a 4\( \pi  \) detector array\cite{Miniball}\cite{Fopi}
can be superior to a finite angle hodoscope, as used in many previous investigations,
provided the granularity (angular resolution), energy threshold, energy resolution
and the particle identification are of sufficient quality. A good 4\( \pi  \)
setup allows minimally biased event selection avoiding many possible distortions.

\textcolor{black}{At GANIL,} \textbf{\textcolor{black}{}}\textcolor{black}{the
system Xe on Sn has been extensively investigated at 50 AMeV using the 4\( \pi  \)
multidetector INDRA \cite{Marie_Xe+Sn2}, \cite{Marie_xe_sn}, \cite{Lukasik}.
Here experimental data are further analyzed by means of correlation functions
aiming at the correlation properties of light particles \cite{DGourio_Thesis}.
They allow to perform event by event the following tasks~: (i)~determine an
'experimental' impact parameter, (ii)~select and characterize the emission sources,
and (iii)~build the correlation functions. Instead of imposing a spatially fixed
particle correlator, we now can use a corr}elator, which fully exploits the
event topology. This correlator continuously adapts to the kinematical configuration
of every single collision. So each emission process can be characterized \textbf{}individually\textbf{.}

This study focuses on two-proton (p-p), two-deuteron (d-d) and proton-deuteron
(p-d) correlation functions for the system \( ^{129}Xe+^{Nat}Sn \) at 50 AMeV
\textcolor{black}{which aims at providing time scales for the emission of the
light charged particles. This should shed light on the type of processes, in
particular if the reaction is} \textbf{\textcolor{black}{}}\textcolor{black}{dominated
by preequilibrium emission} occurring at the early stage of the reaction \textcolor{black}{(typically
with times between 0 and 100 fm/c) or by thermal emission from the projectile
and target like fragments (typically with times of a few hundreds fm/c).} \textcolor{red}{}\textcolor{black}{With
the knowledge on the chronology pattern, the emission time of deuterons might
give a hint to the mechanism for the production of this lowly bound particle
in hot nuclear matter. For this purpose the design of the INDRA detector is
well-suited \cite{Marie_xe_sn}. A large range of excitation energies} (up to
12 AMeV) is covered and the light \textcolor{black}{charged} particle multiplicity
which can be measured is adequate for our selection criteria. 

With a conventional reaction picture in mind, and supported by recent studies
showing that a sizeable fraction of fragments are emitted in the mid-rapidity
region \cite{Lukasik}, only thermalized particles are expected in the forward
hemisphere of the momentum space of the projectile-like source. Such an assumption
has for example important consequences for the estimation of the excitation
energy and the slope parameter of the projectile source \cite{Caloric curve},
\cite{Dissipative Collision}. Conversely semiclassical calculations of heavy
ion reactions in this energy domain have shown that the projectile-like and
the prompt emissions from \textcolor{black}{the interacting zone} present a
large overlap in their rapidity distributions \cite{LV_Eudes}. We have constructed
our correlation functions with particles selected in this forward region to
find out if the thermalized component is really the single contribution.

The extraction of the emission time was performed with a quantum model whose
interesting feature is to take into account the Coulomb effect of the source
charge by analytically solving the three-body problem \cite{FSI1}, \cite{FSI2}.

\section{INDRA setup and light \textcolor{black}{charged} particle resolutions }

The experiment was performed at the GANIL facility where the INDRA detector
has been installed with a target of 350 \( \mu g/cm^{2} \). The beam intensity
was limited to 0.4 nA \textcolor{black}{to avoid a saturation of the data acquisition
software.} 

INDRA \cite{Indra} has been designed to maximize the detection efficiency of
charged particles at intermediate energy. It reaches a total detection efficiency
\textcolor{red}{}\textcolor{black}{of} 90\%. The fine granularity chosen is
such that double counts stay below 5\%. INDRA consists of an array of 336 modules
reparted on 17 rings centered along the beam axis. Each module is a telescope
composed of an ionization chamber (ChIo) filled with \( C_{3}F_{8} \) gas followed
by a Cesium Iodide (CsI) scintillator. For forward angles below 45 degrees,
the resolution is further improved by insertion of a 300 \( \mu m \) silicon
(Si) wafer between the ChIo and the CsI. With \( \Delta E-E \) methods in the
telescope, the charge identification goes up to Z=54. Isotopic resolution (PID)
is obtained for Z=1,2 (and up to Z=5 for ring 2 to 8) by pulse shape analysis
of the CsI light output. Lowest energy threshold for the identification of protons
and deuterons is 6 \textcolor{red}{}\textcolor{black}{MeV} using the matrix
{[}CsI(fast) + Si{]} versus {[}CsI(slow){]}. 

In particular p-p correlation functions require the resolution of very small
relative momenta, less than 20 MeV/c. A minimum relative momentum of 10 MeV/c
can be reached for forward angles under \( 20^{o} \). \textcolor{black}{For
the determination of the particle coordinates, the angle from the target to
the middle of the detector has been used instead of a random distribution over
the spatial extension.} The energy resolution of light particles is between
100 keV and 200 keV depending on the module. With exception of INDRA's first
ring (\( \Delta \Theta =2^{o}-3^{o} \)) which consists of plastic phoswich
detectors (NE102 and NE115) for standing higher particle rates in this region,
the light isotope separation could be performed on the overall domain.

Fig. 1 shows the isotope resolution summed up for rings 2 to 9 (\( 3^{o}<\theta <45^{o} \)).
The insert shows it separately for rings 2, 5 and 9. The projectile-like fragments
at small angles come out at larger energy. To avoid saturation there, the photomultiplier
gains steadily increase from ring 2 to ring 17 \textcolor{red}{}\textcolor{black}{by
about an overall factor of 10.} Subsequently ring 2 has less PID resolution,
seen in the insert to Fig. 1. However the angular resolution of the first rings
is superior and therefore they contribute important information to the correlation
function at small relative momentum.

The p-p correlation function of Fig. 2 shows the data from the forward hemisphere
of the projectile source (FHPS, see section 3) without impact parameter selection.
\textit{}It can be continuously constructed from 10 MeV/c up to 250 MeV/c relative
momenta due to the 4\( \pi  \) coverage and the good angular resolution (yet
the forward source selection slightly increases the minimal relative momentum).
The structure in the correlation function at 20 MeV/c is due to the attractive
s-wave p-p interaction \cite{Koonin}. There is a Coulomb suppression at very
small relative momentum and possibly in the range between 50-75 MeV/c. The normalization
has been applied to the data points between 100 and 120 MeV/c which is above
any remaining two-body effects and below any kinematical effects at higher momentum.
A normalization at lower relative momentum would have introduced a misinterpretation
of the correlation effect.

\section{Event sorting and source selection}

We take in our analysis events in which the total longitudinal momentum of detected
ejectiles is above 80\% of the initial momentum. We refrain from further cuts,
for instance the totally detected charge \( Z_{tot} \), to conserve a representative
impact parameter distribution. For most of the events the target like ejectile
is lost due to \textcolor{black}{the velocity thresholds.} \textcolor{black}{This
missing fragment has been kinematically reconstructed and taken into account
in the calculation of the momentum tensor and thrust variable.} This \textcolor{red}{}\textcolor{black}{event
class g}ives an unbiased starting point for the analysis of light particle correlations.

The calculation of the impact parameter is based on the total transverse energy
of the light charged particles (Z\( \leq  \)2) \textcolor{black}{whose experimental
detection} is quite independent of the reaction mechanism. \textcolor{black}{Furthermore
a recent analysis of the correlation between the total multiplicity versus the
transverse energy (\( E_{T} \)) for that system has demonstrated the validity
of \( E_{T} \) \cite{Lukasik}.}

In order to observe impact parameter dependent properties while maintaining
sufficient statistics, we have defined three \( E_{T} \) bins for which the
correlation function is constructed. The ``peripheral'', ``intermediate\char`\"{}
and ``central'' events have a \( E_{T} \) range between 0-280 MeV, 280-420
MeV and above 420 MeV corresponding to reduced impact parameters in {[}1-0.65{]},
{[}0.65-0.35{]} and smaller than 0.35 respectively.

INDRA allows to build event by event a momentum tensor \cite{Ellipsoide} \textcolor{black}{defined
in the center of mass} by~:

\[
Q_{ij}=\sum ^{M}_{k=1}\frac{1}{p}p_{i}(k)p_{j}(k)\]
 where M is the multiplicity of fragments with a charge Z greater than 2, p
is the momentum of the k'th particle in M and \( p_{i} \), \( p_{j} \) two
of the Cartesian momentum components. The \textcolor{black}{eigenvectors} of
this tensor establish a reference frame. The main axis (eigenvector of the largest
eigenvalue) gives the average direction of nuclear matter emission. The eigenvectors
associated to the two largest eigenvalues define a reaction plane. The FHPS
selections and the calculations have been performed with regards to this new
reference system. Fig. 3 shows the transverse versus the parallel velocity of
the protons when \( E_{T} \) is smaller than 40 MeV. Since this selection implies
only very peripheral events, a clearer separation of the sources is exhibited. 

The next task consists of recognizing the fragments emitted either from the
target-like or from the projectile-like source. The thrust variable defined
by
\[
T=\max _{c_{1},c_{2}}\frac{\mid \sum _{i\in c_{1}}\overrightarrow{P_{i}}\mid +\mid \sum _{j\in c_{2}}\overrightarrow{P_{j}}\mid }{\sum ^{M}_{k=1}|\overrightarrow{P_{k}}\mid }\]

divides up the fragments in two groups \( c_{1} \) and \( c_{2} \) corresponding
to the two emitting sources. The velocity of each of them is determined by a
kinematic reconstruction within these two ensembles \cite{Metivier_Thrust}.
As an example, the average velocity of the projectile-like source is drawn with
a vertical bold line in Fig. 3.

To disentangle the projectile-like source emission from mid-rapidity contributions,
we have taken particles with a parallel velocity larger than the projectile
source velocity. This region \label{FHPS}(FHPS) is on the right of the bold
line in Fig. 3. \textcolor{black}{In the following, this sample of particles
is used as the base for the extraction of the slope parameter in the energy
spectra as well as for the construction of the correlation functions. We remind
the reader here that our correlator works dynamically within the reference frame,
a fact being imperative in order to optimize the FHPS selection.}

\section{Source parameters : disentangle size and time}

\subsection{\textit{Size} }

The shape and the height of the correlation function is given by the strength
of the interactions which themselves depend on the average distance between
the two particles detected in coincidence. This distance depends on the average
emission time and on the spatial source extension. The double parametrization
can be circumvented only for two extreme cases: at high energy the emission
time is set to zero while at low energy the emission time is very long and consequently
the source size is negligible \cite{Fixed_source_size}. In our energy domain,
both parameters are relevant. Using the complete detection by INDRA of all charged
products we alternatively can determine the source size directly. 

\textcolor{black}{For this estimation, ejectiles are grouped according to (\( Z\leq Z_{L} \))
and (\( Z>Z_{L} \)) where \( Z_{L} \) is an adjusting parameter. In the first
group only the particles faster than the projectile-like source velocity are
included (particles from the FHPS) and the sum of their charges is multiplied
by 2 taking into account the isotropical projectile-like emission. This sum
is \( Z_{P} \) in which the additional mid-rapidity contribution mainly composed
of light charges is suppressed. The second group contains only the particles
faster than the center of mass velocity, the sum of their charges is \( Z_{F} \).
This separate treatment of the heaviest particles (second group) takes into
account an asymmetry of emission in the projectile-like reference frame (for
example only one big remnant, or two fission fragments). The total charge of
the source is \( Z_{Total}=Z_{P}+Z_{F} \). To test the quality of this procedure,
the calculation of \( Z_{Total} \) has been performed for different values
of the parameter \( Z_{L} \)(2, 4, 8, 10). Fig. 4 shows that \( Z_{Total} \)
only varies by less than 6\% with \( Z_{L} \). For later calculation of \( Z_{Total} \)
we chose \( Z_{L}=4 \).}

\textcolor{black}{For each of the three impact parameter bins we have also defined
\( Z_{Max} \) as the largest fragment being faster than the center of mass.
\( Z_{Max} \) decreases with the centrality as expected in the geometrical,
simple picture of the collision also shown in Fig. 4. \( Z_{Total} \) unexpectedly
remains constant. }

\textcolor{black}{To estimate the source size, we make two assumptions about
the projectile-like source. First, the \( A_{Total}/Z_{Total} \) ratio is fixed
to the one from the valley of nuclear stability. Secondly, with this value of
\( A_{Total} \) we determine the source radius r by assuming a normal nuclear
density \( \rho _{0} \) and by simply applying \( r=r_{0}.A^{1/3}_{Total} \)
with \( r_{0}=1.2 \) fm. The central 'single source' events which have a radial
flow of 2 AMeV \cite{Radial_flow_3} have a very small cross section. This implies
that a density different from \( \rho _{0} \) does not make sense for our calculation
even in the central impact parameter bin. The extracted source sizes are given
in Table 1. The errors are derivated from the \( Z_{Total} \) distribution
widths.}

\subsection{\textit{Slope parameter}}

\textcolor{black}{We have extracted the} \textit{\textcolor{black}{slope parameter}}
\textcolor{black}{from the experimental energy spectra using the formula for
surface emission. As for the source size estimation, we have selected only the
protons and the deuterons located in the FHPS region. Their kinetic energy is
given in the projectile-like source reference frame. As example for the peripheral
collisions, Fig. 5 shows the energy spectra restricted to ring 2, 4 and 6 where
the double slope is the most clearly visible. The low energy one originates
from the projectile-like thermal emission, the other at higher energy, presumably
from a preequilibrium emission. We observe that the importance of high energy
particles decreases with the radial angle as if this emission were concentrated
along the beam axis, and as expected in a Fermi-jet picture \cite{Fermi-jet}.
The slope parameters averaged over all the rings included in FHPS are given
in Table 1.} \textcolor{black}{For the quantum calculation code} \textbf{\textcolor{black}{}}\textcolor{black}{presented
below we have used a weighted average value of the double slopes. }

\section{Description of the quantum model}

The extraction of the emission time has been performed by using the three-body
quantum model developed by R. Lednicky \cite{FSI1}, \cite{FSI2}. This code
calculates the quantum statistics for identical particles and the final state
interaction by taking into account the nuclear and the Coulomb potentials. The
Coulomb repulsion on the particle pair due to the emitting projectile-like source
is also included \cite{Martin_3Body}. The three-body problem is analytically
solved by making an adiabatic assumption~: the relative motion between the two
particles has to be much slower than their velocity in the source reference
frame. \textcolor{red}{}\textcolor{black}{Due to a sizeable angle between neighboring
detectors, the small relative momentum region can only be populated by pairs
of particles with almost equal velocity.} Thus the adiabatic assumption \textcolor{black}{is
fulfilled in the region of the signal.}

The introduction of the emitter Coulomb effect in the quantum calculation is
a new feature brought by this model. We feel that this type of description is
required to correctly reproduce the experimental data because the presence of
the remnant source charge is intrinsic to the model. In so far it may surpass
models where the Coulomb influence of the emitter is only treated as a correction
\cite{Gamov_factor}. \textcolor{black}{Limits of this model are certainly related
to the static source description \cite{Optimized_Source} which does not take
into account the correlation between momentum and position, nor the dynamical
emission. }

\textcolor{black}{The particle pairs are generated by a static surface emitter
describing the projectile-like source with size and slope parameter (see Table
1). The choice of a surface emission instead of a volume one is justified on
grounds of the dominance of binary processes \cite{Metivier_Thrust}, \cite{Binary_Dominance}
meaning a big fragment at the speed of the projectile remains in the exit channel.
Combined with the fact that no radial flow is observed (see 4.1), the picture
of a surface emission of particles seems to be the most appropriate.} The time
distribution for particle emission follows an exponential decay law with average
emission times to be determined from comparison with the data. The energy distribution
is taken from the experimental slope parameter. The distributions have been
filtered according to the geometrical granularity, including also double counts.
In addition, energy and particle identification thresholds as well as the source
selection criteria (FHPS) have been folded in.

\section{Analysis of the light particle correlation functions}

The experimental correlation function is defined by the yield of true coincidences
as a function of the two particle relative momentum divided by the so-called
false coincidences obtained by the technique of event mixing \cite{Mixing}.
It consists of taking two particles from two different events which assures
a full decorrelation and has the advantage to use the same sample of events
for true and false coincidences. Possible distortions coming from our FHPS selection
are then largely eliminated. In general, event mixing introduces an additional
term in the relative momentum due to different source velocities. This influence
which otherwise would disturb the correlation function is reduced by the impact
parameter selection.

\subsection{The Proton-Proton correlation function}

The p-p correlation functions for the three intervals in impact parameter are
represented in Fig. \textbf{}6. \textbf{}The statistics of 2.3 millions of reconstructed
events give a reasonable correlation function above 10 MeV/c relative momentum.
In all three event classes, the resonance at 20 MeV/c is clearly visible and
well described by a time value of \( \tau  \)=80 fm/c \textcolor{black}{with
a surface emission model.} A time variation of 25\% changes the height of the
resonance by a factor of two. This demonstrates the high sensitivity of the
correlation function on the emission time parameter. \textcolor{black}{The fact
that the time estimation strongly depends on the emission description has also
been investigated~: a simulation using volume break-up, closer to the scenario
of a preequilibrium emission, leads to shorter times in the range of 25 fm/c.}

\textcolor{black}{The undershoot in the shape of the calculated correlation
function visible at 50 MeV/c is induced by the Coulomb repulsion between the
two particles. The additional boost due to the repelling charge of the third
body (emitting source) shifts the Coulomb suppression, which is usually located
at small relative momentum, to higher values. In a pure two body calculation
this undershoot almost disappears.}

\textcolor{black}{The experimental data also show this behavior at 50 MeV/c}
\textbf{\textcolor{black}{}}\textcolor{black}{for the central events contributing
to a very good agreement with the calculation. On the contrary the experimental
correlation function for the peripheral events does not show this anticorrelation
feature. It is not fully reproduced by the calculation based still on a surface
emission. This might indicate the presence of a preequilibrium component, for
which no coherent Coulomb influence of a source is expected and for which a
volume break-up simulation would be more appropriate. A recent QMD calculation
for the system Xe+Sn at 50 AMeV \cite{Time_Scale_Nebauer} predicts the compression
to be maximal at 50 fm/c after the beginning of the collision compatible with
the total time of spatial overlap (\( \simeq  \) 40 fm/c). The end of fragment
interaction occurs in this model at 120 fm/c considered as the end of the reaction.
Taking this time scale into account, our measured times in between 25 and 80
fm/c are compatible with a preequilibrium emission. It is confirmed by the disappearance
of the Coulomb undershoot at 50 MeV/c in the case of the peripheral collisions.
The appearance of the double slope in the energy spectra supports this interpretation.}

One must ask here, how the emission time alters for a given error in source
size. To test this we have reduced the charge from Z=46 to Z=36, \textcolor{black}{simply}
assuming all the Z=1 particles do not belong to the projectile-like emission.
The result for the emission time is then 100 fm/c instead of 80 fm/c which still
stays short enough to be compatible with our conclusion saying, these light
particles characterize a prompt process of pre-equilibrium emission which covers
a large domain of rapidity. \textcolor{black}{By lack of statistics resulting
from our source and impact selections, the method of simultaneous determination
of the source size and the emission time from parallel and transverse correlation
functions \cite{pp_psi_lisa} could not be tested here.}

\subsection{The Deuteron-Deuteron correlation function}

The d-d correlation function has been constructed on the same base of events
as p-p. \textcolor{red}{}\textcolor{black}{The normalization has been applied
to the data points between 150 and 200 MeV/c. Again the three impact parameter
classes have been separately analyzed and} the results are shown on Fig. \textbf{\large }7.
We immediately observe the anticorrelation effect in the d-d correlation function
for small relative momenta. Despite the fact that data do not go below 30 MeV/c
the fit of the quantum model gives the following results : for peripheral reactions
the emission time is as least 200 fm/c, for semi-central reactions it is 100
fm/c and for central it is 25 fm/c. This behavior can be interpreted as an increasing
contribution of out-of-equilibrium emission.

Yet the creation of barely bound particles is not the prefered mechanism of
hot nuclear matter to dissipate energy. \textcolor{black}{A second scenario
could be imagined, assuming the deuteron creation happens only at a certain
low density \cite{Low_density_for_deuteron}. Then the extracted emission time
would give a direct hint on when this state of the nuclear matter is reached
during the reaction process. However the double slope of the deuteron energy
spectra seems to favor the} out-of-equilibrium emission. In the picture of the
coalescence model \cite{Coalescnence_Deuteron}, the deuteron formation is directly
connected to the proton creation. Consequently it is not surprising to find
also two components in the deuteron energy spectra.

In conclusion the process of deuteron production remains an open question. Still
we tend to favor the preequilibrium emission over the other explanations. Unfortunately\textcolor{black}{,
the present sample of data is insu}fficient to disentangle more.

\subsection{The Proton-Deuteron correlation function and the emission chronology}

The correlation function of non-identical particles can give model independent
information about their mean order of emission simply making velocity selections
\cite{Order_lednicky}, \cite{Order_gelderlos}, \cite{Order_Soff}. We have
applied this method to the p-d correlation function for particles emitted in
the FHPS region. The principle is to compare two functions. The first, ( \( 1+R^{+} \)),
is constructed with pairs where the proton is faster than the deuteron in the
projectile-like source reference frame. The second function, ( \( 1+R^{-} \)),
corresponds to the reverse situation . When the first emitted particle is slower
than the second, the average distance will be reduced and the Coulomb suppression
effect enhanced, and vice versa. The comparison of the two functions gives the
mean order of emission as it is shown in Fig. 8 for the peripheral collisions.
The Coulomb suppression is more pronounced in \( 1+R^{+} \), which the ratio
clearly demonstrates. This means that the deuteron is on average emitted earlier
than the proton, namely \( \tau _{deuteron}<\tau _{proton} \). The same time
sequence is observed for the two other impact parameter selections. \textcolor{red}{}\textcolor{black}{It
is important to note that} this result is only validated between 0-120 MeV/c
relative momentum where the anticorrelation effect leaves a measurable signal. 

The chronology of emission in p-d spectra of peripheral collision \textcolor{black}{can
be considerated} as surprising since we just learned from previous paragraphs
that the mean emission time is 80 fm/c for p-p and 200 fm/c for d-d. This apparent
contradiction can be resolved by postulating that the protons which contribute
to p-p are not identical with those contributing to p-d. Indeed the proton energy
spectra show a fast and a slow component (Fig. 5)\textcolor{black}{. Furthermore,}
from the \textcolor{black}{energy slopes of Fig. 9, we infer that the protons
of p-p in the relative momentum range of 0-50 MeV/c have a higher mean kinetic
energy (12.9 MeV) than the one of p-d in the same range (9.9 MeV). Subsequently
it is clear that the p-p correlation function is more influenced by the preequilibrium
protons. The extracted times between 25 and 80 fm/c must be seen as an upper
limit reflecting the mixing of a fast and a slow component. In the p-d correlation
function the situation is different~: the coincidences of ''early'' protons
with the deuterons are shifted to higher relative momentum value since the speed
differs more than for thermal proton-deuteron pairs. Moreover, the functions
\( 1+R^{+} \) and \( 1+R^{-} \) are built in the projectile-like source frame.
So protons from the interacting zone which feed both correlation functions are
per definition faster. Consequently they put only little weight to the emission
time order determination. The p-d correlation function informs rather about
the thermal protons }

\textcolor{black}{Taking the deuteron emission time extracted from d-d, \( \tau = \)
310, 110 and 25 fm/c for the peripheral, intermediate and central events respectively,
we have deduced the corresponding time for the protons (Fig. 10) and obtained
400 fm/c for central collisions. Unfortunately n}o satisfying agreement can
be found between the quantum model and the p-d data for peripheral and int\textcolor{black}{ermediate
reactions, possibly due to the different proton contributions, one acting at
low the other at high relative momentum. In addition, the description of the
source without dynamical features might prevent a better agreement. Therefore
the extracted parameters do not retain the meaning of a physical time.}

Although both components of protons simultaneously play a role in the correlation
function, we can deduce a chronology pattern of the light particle emission.
The fast protons from the interacting zone come first, then the deuterons and
still later the protons thermally emitted by the projectile-like source. It
is possible to better separate the two proton components by making a cut on
the parallel velocity as Fig. 11 shows. The clear enhancement of the p-p resonance
at 20 MeV/c and the disappearance of the Coulomb suppression at 50 MeV/c reveal
the enlarged part of fast protons not feeling \textcolor{black}{the charge of
an emitter. The importance of differentiating between short-lived and long-lived
emission components and subsequent space-momentum correlations has also been
discussed by using source imaging methods} \textbf{\textcolor{black}{\cite{Optimized_Source}.}}
\textcolor{black}{Furthermore, making the plausible hypothesis that fastest
particles are emitted earlier than the slowest ones because the available energy
is greater at the beginning of the reaction \cite{He3_He4_Time}, we can extrapolate
from the slope parameter pattern an equivalent chronology~:} \textbf{\textcolor{black}{
\[
T_{Proton\, 2}>T_{Deuteron\, 2},\: T_{Deuteron\, 1}>T_{Proton\, 1}\]
}}

\section{Conclusion}

We have taken the large data set for collisions of Xe on Sn at 50 AMeV which
INDRA has accumulated at GANIL to examine correlations of protons and deuterons.
Such studies have many attractive aspects in view of the complete detection
of all collision residues by a 4\( \pi  \) detector. Foremost to name is the
unique possibility to well determine the emitting source in particular for symmetric
systems at intermediate energy. In addition each event can be individually characterized
in its own frame of reference, due to full charge, angular and energy coverage
of INDRA. 

The time scale as well as the chronology of emission of light projectile-like
particles could be determined from two particles correlation functions interpreted
by a full three body quantum code. Total charge measurements as a function of
impact parameter indicate possible out-of-equilibrium emission of protons from
the forward hemisphere of the projectile source. The observation that the slope
parameter of the energy spectra exhibits two components points to a similar
conclusion. Strong confirmation of these findings stems from the very short
emission time extracted from p-p correlation functions. While this process is
expected to dominate in central collisions our study unveils that it also contributes
to the forward zone of binary peripheral collisions. We explained in the text
how both proton components are not equally shared in p-p and p-d correlation
functions. So the whole emission chronology pattern remains self-consistent.
It is also in good agreement with the measurement of slope parameters. The short
time scale in p-p reveals the presence of fast hot protons from an out-of-equilibrium
process. Protons emitted later than the deuterons correspond to the really equilibrated
production from the projectile-like source. The light particle emission chronology
including the deuteron formation via the \( NNN\rightarrow dN \) process has
been calculated in the Boltzmann-Uehling-Uhlenbeck (BUU) approach \cite{Order_BUU}.
The theoretical results although for lighter systems are in excellent agreement
with the present experimental study. 

We would have liked to look into the hydrogenic correlation function in more
detail by finer selecting impact parameter intervals. Furthermore, the inclusion
of tritons could have given valuable additional information. This task cannot
yet be performed on the same footing as with protons and deuterons due to a
serious shortage of data statistics.

We therefore suggest to perform a high statistics experiment especially dedicated
to light particle correlations. INDRA parameters, optimized toward this goal,
could contribute important and still better information on the dynamics of light
particle emission.

\section{Acknowledgments}

The authors whish to thank H.Orth for his careful reading of the manuscript,
R.Lednicky for discussions and providing his code. D.G acknowledges the support
of the ALADIN group at GSI.
\newpage

\vspace{0.3cm}
{\centering \begin{tabular}{|c|c|c|c|c|c|}
\hline 
\multicolumn{2}{|c|}{}&
\multicolumn{2}{|c|}{T \textbf{\footnotesize \( _{Proton} \) (MeV)}}&
\multicolumn{2}{|c|}{ T \textbf{\footnotesize \( _{Deuteron} \)(MeV)}}\\
\hline 
\textbf{\footnotesize Centrality}&
\textbf{\footnotesize Radius (fm)}&
\textbf{\footnotesize T}\( _{1} \)&
\textbf{\footnotesize T}\( _{2} \)&
\textbf{\footnotesize T}\( _{1} \)&
\textbf{\footnotesize T}\( _{2} \)\\
\hline 
\hline 
\textbf{\footnotesize Peripheral}&
\textbf{\footnotesize 5.9\( \pm  \)0.3}&
\textbf{\footnotesize 3.7\( \pm  \)0.1}&
\textbf{\footnotesize 9.5\( \pm  \)0.1}&
\textbf{\footnotesize 4.1\( \pm  \)0.1}&
\textbf{\footnotesize 8.9\( \pm  \)0.1}\\
\hline 
\textbf{\footnotesize Intermediate}&
\textbf{\footnotesize 5.9\( \pm  \)0.4}&
\textbf{\footnotesize 4.8\( \pm  \)0.1}&
\textbf{\footnotesize 11.3\( \pm  \)0.1}&
\textbf{\footnotesize 6.0\( \pm  \)0.1}&
\textbf{\footnotesize 10.7\( \pm  \)0.1}\\
\hline 
\textbf{\footnotesize Central}&
\textbf{\footnotesize 5.9\( \pm  \)0.5}&
\textbf{\footnotesize 5.2\( \pm  \)0.1}&
\textbf{\footnotesize 12.8\( \pm  \)0.1}&
\textbf{\footnotesize 11.3\( \pm  \)0.1}&
\textbf{\footnotesize {*})}\\
\hline 
\end{tabular}\footnotesize \par}
\vspace{0.3cm}

Table \textbf{1} : Parameters of the projectile-like source. The radius has
been calculated from the total reconstructed charge of the source, by assuming
a A/Z ratio in the valley of nuclear stability and a normal nuclear density.
The slope parameter has been extracted from the energy spectra in the source
reference frame. 

{*}) Only one slope could be extracted.

\vspace{0.3cm}
{\centering \includegraphics{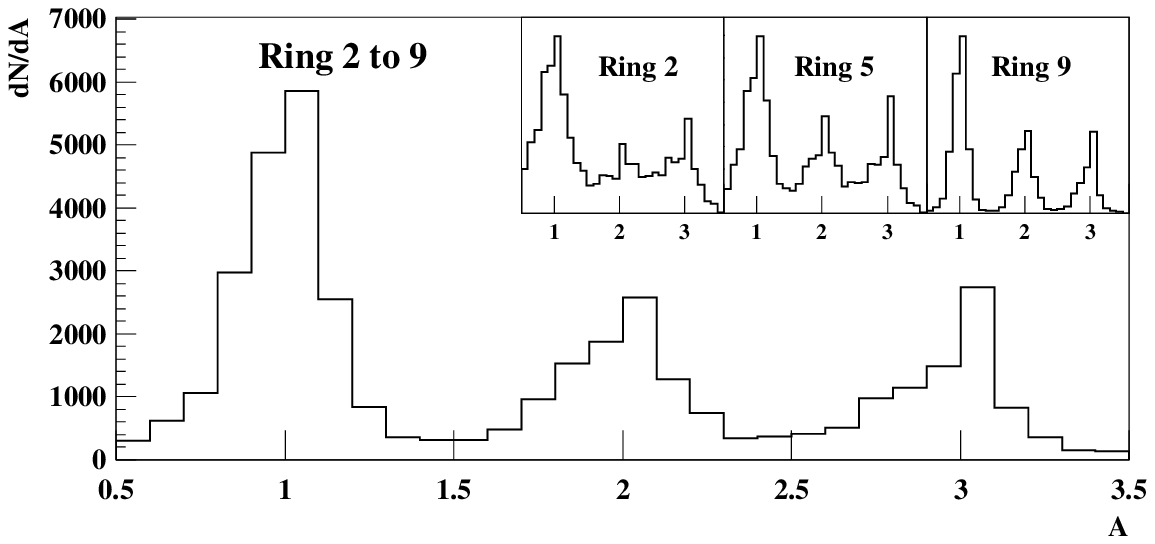} \par}
\vspace{0.3cm}

Fig. 1 : Light isotopes resolution (Z=1) of the 8 first rings and in the insert
for the rings 2, 5 and 9 separately. 

\vspace{0.3cm}
{\centering \includegraphics{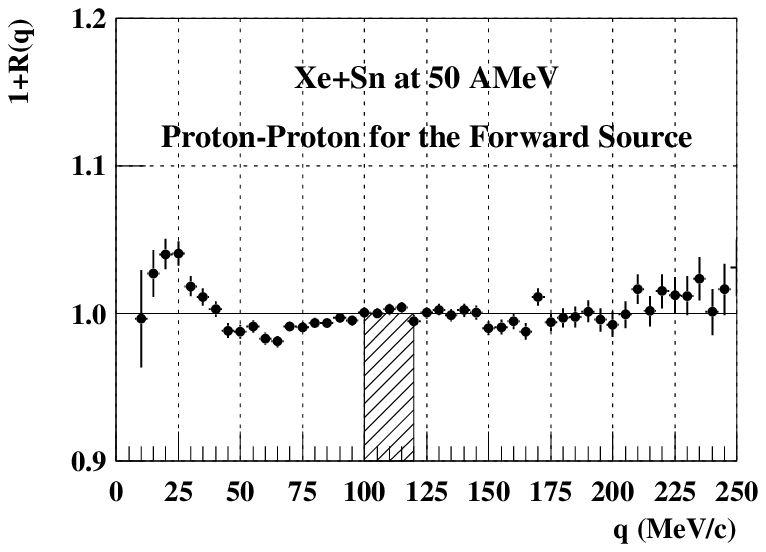} \par}
\vspace{0.3cm}

Fig. 2 : Two-proton correlation function for the FHPS selection.

\vspace{0.3cm}
{\centering \includegraphics{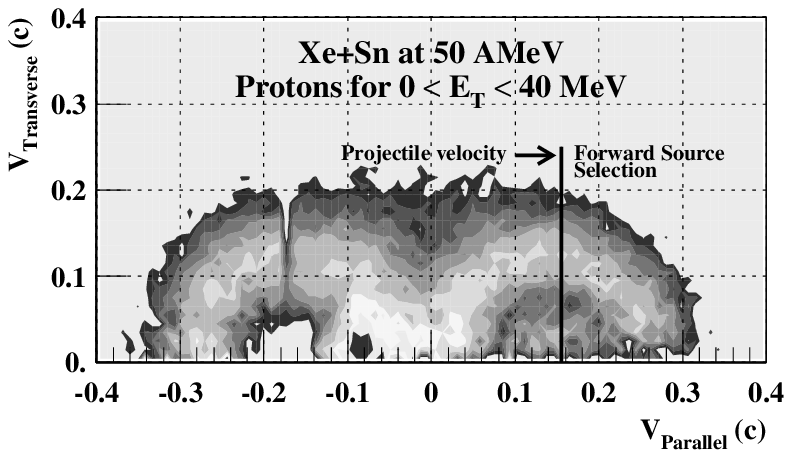} \par}
\vspace{0.3cm}

Fig. 3 : Invariant velocity plot of the protons in the center of mass for a
total light particles transverse energy smaller than 40 MeV which corresponds
to a normalized impact parameter larger than 0.9. The average value of the reconstructed
forward source is represented by a vertical line at \( V_{//}=0.155c \). This
line position is in accordance to the middle of the Coulomb circle. The FHPS
is defined by the particles in each event which are faster than the reconstructed
forward source velocity.

\vspace{0.3cm}
{\centering \includegraphics{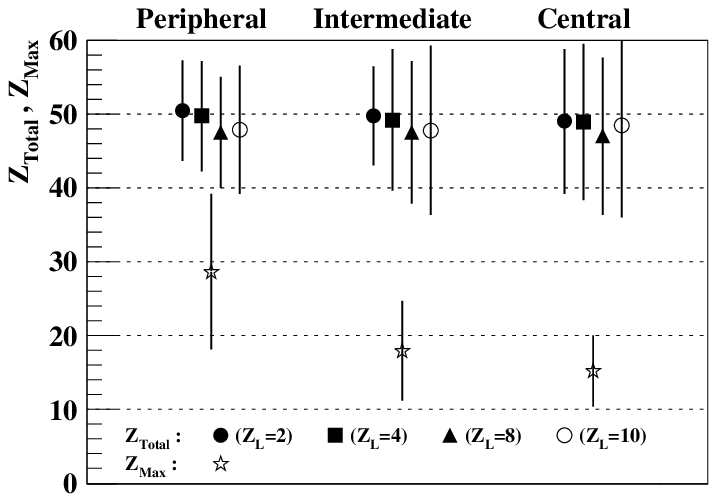} \par}
\vspace{0.3cm}

\textcolor{black}{Fig. 4 : The reconstructed total charge of the projectile-like
\( Z_{Total} \) for three centrality bins and for different calculations. The
meaning of the parameter \( Z_{L} \) is given in the text. The \( Z_{Total} \)
dependance on \( Z_{L} \)is less than 10\% which places confidence in this
estimation. The largest fragment \( Z_{Max} \) (open stars) gets smaller with
centrality as expected in a geometrical picture. This is not the case of \( Z_{Total} \)
which remains constant.}

\vspace{0.3cm}
{\centering \includegraphics{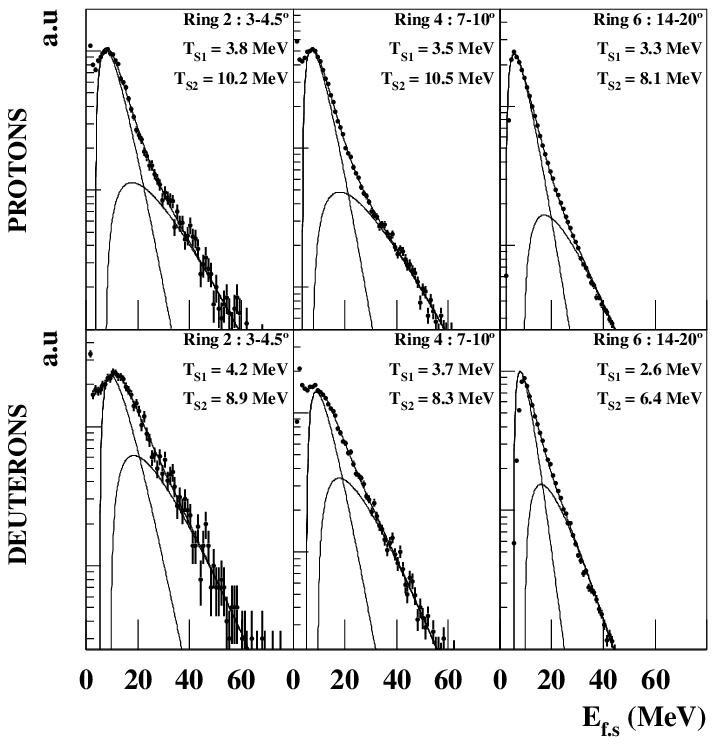} \par}
\vspace{0.3cm}

Fig. 5 : The energy spectra of the protons (top) and deuterons (bottom) for
the ring 2, 4 and 6 in the case of the peripheral collisions. The shapes clearly
exhibit two components, better separated in the case of the protons. For comparison
reasons the relative scale is the same for all panels.

\vspace{0.3cm}
{\centering \includegraphics{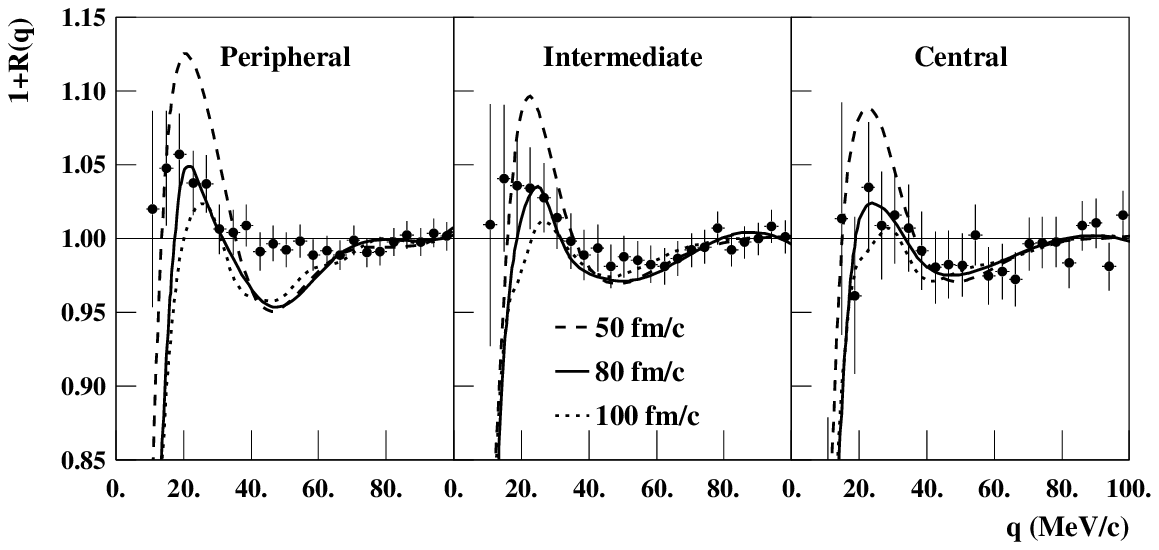} \par}
\vspace{0.3cm}

Fig. 6 : The experimental p-p correlation functions (black circles) of the projectile-like
source for the three impact parameter intervals. Each case has been calculated
with different emission times using the source parameters of Table 1. A time
of 80 fm/c was found to be the best for all impact parameter bins. The resonance
of the calculated function decreases with the centrality because the slope parameter
of the emitter increases.

\vspace{0.3cm}
{\centering \includegraphics{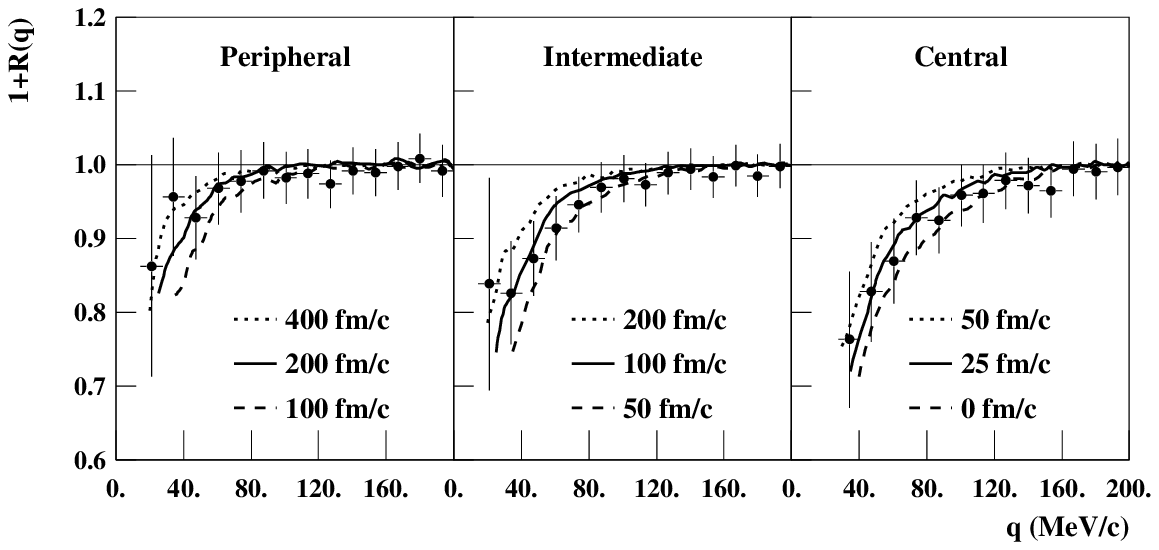} \par}
\vspace{0.3cm}

Fig. {\large }7 : The experimental d-d correlation functions of the projectile-like
source for the three impact parameter bins (black circles). Each case has been
calculated with different emission times using the source parameters of Table
1. There is only a weak dependence on the long emission time parameter for the
peripheral reactions.

\vspace{0.3cm}
{\centering \includegraphics{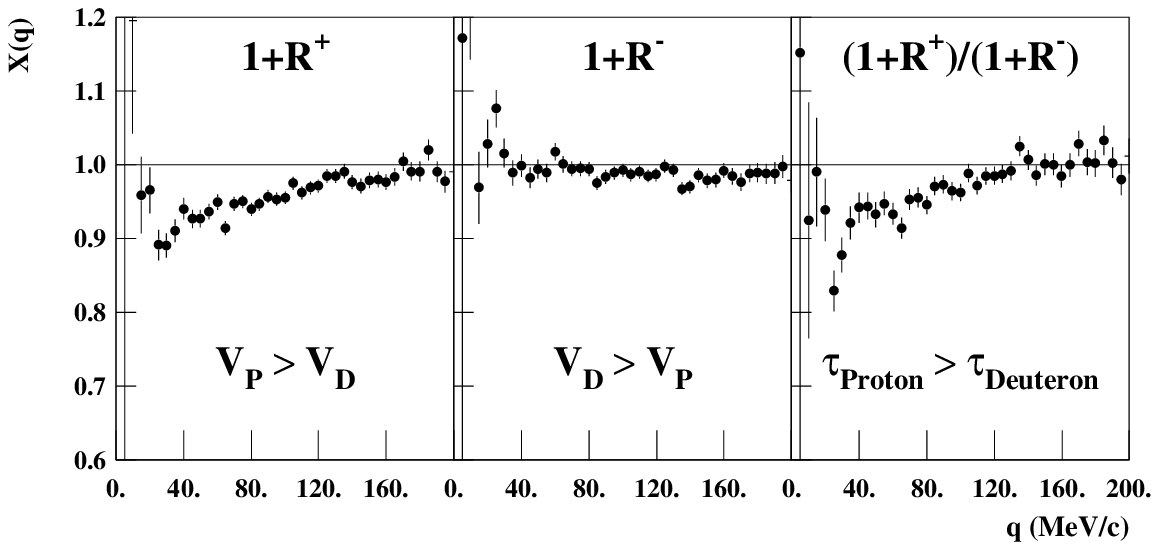} \par}
\vspace{0.3cm}

Fig. \textbf{\large }8 \textbf{\large }: The experimental proton-deuteron correlation
functions of the projectile-like source for the peripheral collisions grouped
into two velocity bins (left and middle panel). \( 1+R^{+} \) contains all
the pairs of particles where the proton was faster than the deuteron. \( 1+R^{-} \)
is the reverse situation. The ratio of both functions (right panel) which is
smaller than unity indicates that the deuteron is on the average emitted earlier
than the proton.

\vspace{0.3cm}
{\centering \includegraphics{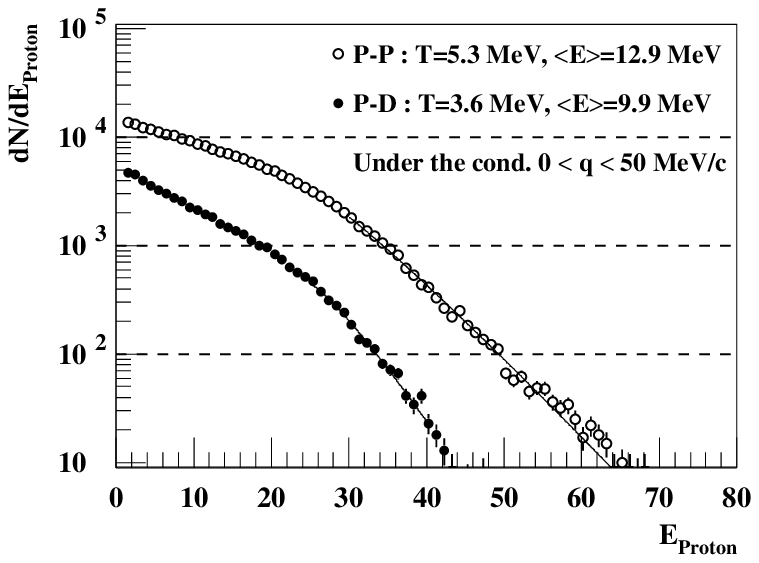} \par}
\vspace{0.3cm}

Fig.9 : The kinetic energy of the protons in the projectile-like source frame
which contribute to p-p (open circle) or to p-d (black circle) with a relative
momentum selection smaller than 50 MeV/c for peripheral collisions. For p-p
the average energy and the slope parameter are higher than for p-d.

{\centering \includegraphics{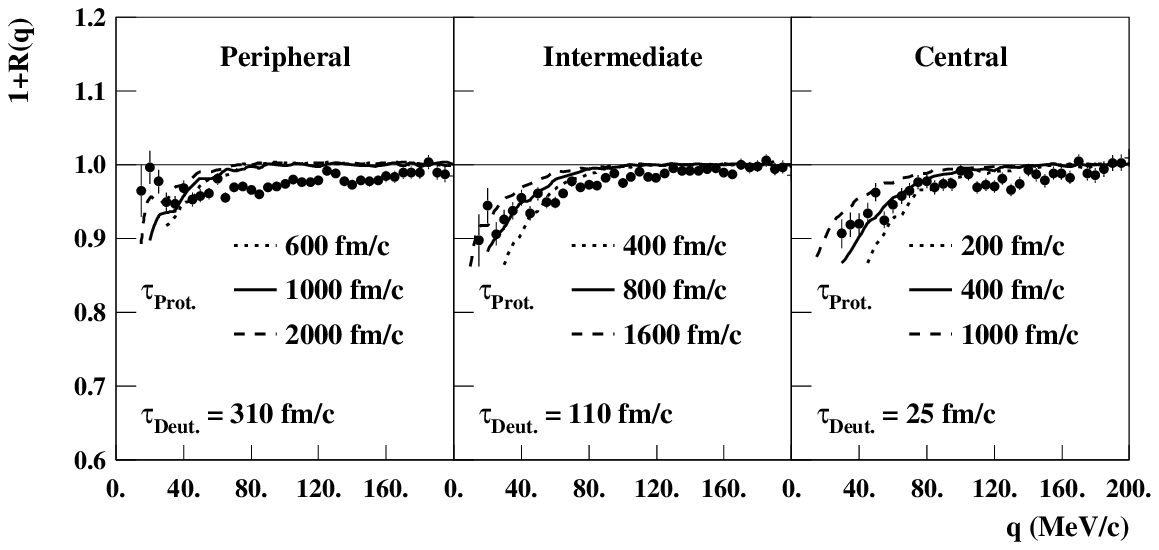} \par}

Fig.10 : The experimental proton-deuteron correlation functions of the projectile-like
source for the three impact parameter selections (black circles). The calculation
(lines) fails to reproduce the data for the peripheral and the intermediate
collisions which may reveal the double contribution of fast and slow protons
in the interference pattern.

\vspace{0.3cm}
{\centering \includegraphics{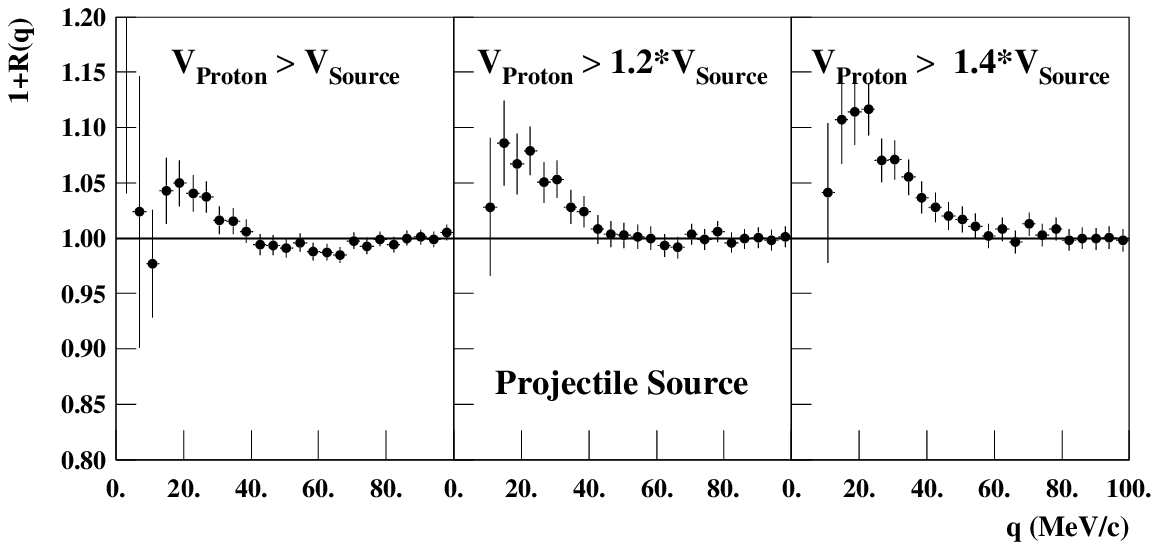} \par}
\vspace{0.3cm}

Fig. 11 : The experimental proton-proton correlation functions for all impact
parameters with increasing selection on their longitudinal velocity in the projectile-like
reference frame. The higher resonance indicates a faster emission time.

\end{document}